\documentclass[prl,reprint,superscriptaddress,showpacs,preprintnumbers,nobibnotes,amsmath,amssymb, aps,longbibliography]{revtex4-2}

\usepackage{cancel}
\usepackage{accents}
\usepackage{mciteplus,slashed}
\usepackage{amssymb,cancel,amsmath}
\usepackage{dcolumn}
\usepackage{bm}
\usepackage[caption=false]{subfig}
\usepackage{appendix}
\usepackage{feynmp-auto}
\usepackage[T1]{fontenc}	
\usepackage{csvsimple}
\usepackage[colorlinks=true,citecolor=blue,linkcolor=blue]{hyperref}
\usepackage[section]{placeins}
\usepackage[capitalise]{cleveref}
\usepackage{booktabs}
\usepackage{graphicx}
\usepackage{subfig}
\usepackage{mathrsfs}
\usepackage{multirow}
\usepackage[utf8]{inputenc}
\usepackage{siunitx}
\usepackage{comment}
\usepackage{gensymb}
\usepackage{enumitem}

\usepackage[dvipsnames]{xcolor}
\usepackage[normalem]{ulem}

\begin{document}
\preprint{RIKEN-iTHEMS-Report-25}
\title{Explosive production of Higgs particles and implications for heavy dark matter}
\author{Seishi Enomoto}
\affiliation{School of Physics, Sun Yat-sen University, Guangzhou 510275, China}
\affiliation{Department of Physics, Faculty of Engineering Science, Yokohama National University, 
Yokohama 240-8501, Japan}
\affiliation{Department of Science and Engineering, Faculty of Electrical Engineering, 
Kyushu Sangyo University, Fukuoka 813-8503, Japan}
\author{Nagisa Hiroshima}
\affiliation{Department of Physics, Faculty of Engineering Science, Yokohama National University, 
Yokohama 240-8501, Japan}
\affiliation{RIKEN Center for Interdisciplinary Theoretical and Mathematical Sciences(iTHEMS), 
RIKEN, Wako 351-0198, Japan}
\author{Kohta Murase}
\affiliation{Department of Physics, The Pennsylvania State University, University Park, 
Pennsylvania 16802, USA}
\affiliation{Department of Astronomy and Astrophysics, The Pennsylvania State University, 
University Park, Pennsylvania 16802, USA}
\affiliation{Center for Multimessenger Astrophysics, The Pennsylvania State University, 
University Park, Pennsylvania 16802, USA}
\author{Masato Yamanaka}
\affiliation{Department of Physics, Faculty of Engineering Science, 
Yokohama National University, Yokohama 240-8501, Japan}
\affiliation{Department of Advanced Sciences, Faculty of Science and Engineering, 
Hosei University, Tokyo 184-8584, Japan}
\affiliation{Department of Literature, Faculty of Literature, 
Shikoku Gakuin University, Kagawa 765-8505, Japan}

\begin{abstract}
It is widely believed that the parameter space for Higgs-portal dark matter that achieves the 
relic abundance through thermal freeze-out has already been tightly constrained, typically 
at masses on the order of ${\cal O}(10-100)$~GeV. We point out the possibility that the 
multiple Higgs production due to its self-interaction dramatically changes this picture. 
We show that the multiplicity can be as large as ${\cal O}(200)$ for the parameters of the 
Standard Model Higgs, independently of the kinematics of the particle production process. 
Consequently, heavy Higgs-portal dark matter of $m_\chi\gtrsim{\cal O}(1)$~TeV can 
achieve the required relic abundance in the same mechanism with that for canonical 
weakly interacting massive particle models.
\end{abstract}

\date{\today}

\maketitle

{\it Introduction---}%
The nature of dark matter (DM) is a long-standing puzzle in 
particle physics and cosmology~(e.g.,~\cite{Bertone:2004pz,Feng:2010gw}). 
The most appealing framework to generate DM is the thermal freeze-out mechanism, 
which links the DM relic density and the properties of particle DM 
independently of the initial conditions in the early Universe. 
Within this framework, carrying out a detailed computation of the relic 
density comes to understand the DM nature. 

The discovery of the Higgs boson has opened new doors for exploring fundamental physics beyond the standard model (SM), such as the nature of DM. 
There are still mysteries to be unraveled in the Higgs sector, such as the origin of the electroweak symmetry breaking and the number of Higgs fields. 
Unified pictures for describing the nature of Higgs fields and DM have been proposed and 
argued in a large amount of literature~\cite{Silveira:1985rk,McDonald:1993ex,Burgess:2000yq,Davoudiasl:2004be,Patt:2006fw,Andreas:2008xy,Andreas:2010dz,Englert:2011yb,Lebedev:2011iq,Chu:2011be,Kanemura:2010sh,Lopez-Honorez:2012tov,Djouadi:2011aa,Baek:2012se,Djouadi:2012zc,Greljo:2013wja,Buckley:2014fba}. An attractive scenario among them is the so-called Higgs portal DM scenario, in which the SM Higgs field acts as a bridge between the fundamental theory behind DM and our world. 
The observed DM abundance in these scenarios is achieved through the resonant 
annihilation of DM to the SM Higgs boson, and then the DM mass is required to be 
$\simeq m_H/2$, where $m_H \simeq 125$\,GeV is the Higgs boson mass~\cite{Arcadi:2021mag}.

The possibility of the {\it Higgsplosion},
which is an exponential growth of the decay rate of highly-excited Higgs 
boson into multiple Higgs bosons, is pointed out in Refs.~\cite{Khoze:2017tjt, 
Khoze:2017ifq, Khoze:2022fbf, Jaeckel:2018tdj, Khoze:2018mey}: 
This cannot be tested with the current facilities 
due to the kinematical limitation. However, it would be possible with future 
collider experiments~\cite{Degrande:2016oan,Gainer:2017jkp,Khoze:2017uga}. 
Studies of factorial growth of the multiparticle production amplitude 
have a long history~\cite{Libanov:1994ug, Libanov:1995gh, Son:1995wz, Makeenko:1994pw}. 
The production cross sections show a characteristic 
exponential form, $\sigma_n \propto \exp[n F(g,n)]$, where $F(g,n)$ is 
a function of the coupling $g$ and the multiplicity $n$. 
Although the naive consideration about the Higgsplosion may result in the continuous 
growth of multiparticle production as the energy scale gets higher, the perturbative 
approach breaks down when the theory becomes strongly coupled and more than 
${\cal O}(100)$ Higgs bosons are produced~\cite{Degrande:2016oan, Khoze:2017ifq, 
Khoze:2017tjt}. At such high energies, the dressed propagator plays an important role 
in preserving the unitarity and regulating the multiparticle production, which is 
called the Higgspersion effect. Details about the competition between the Higgsplosion 
and Higgspersion are not yet fully investigated.

This Letter focuses on the explosive Higgs production in DM annihilation, incorporating 
the effects from the Higgspersion, then 
addresses its phenomenological consequences, such as the DM freeze-out 
in the early universe~
\footnote{Although both the nonperturbative 
aspects and the effect of electroweak symmetry restoration for the 
multiparticle processes are still under debate~\cite{Voloshin:2017flq, 
Demidov:2022ljh, Curko:2019dtu, Schenk:2021yea, Abu-Ajamieh:2020wqn, 
Monin:2018cbi, Belyaev:2018mtd, Demidov:2018czx, Demidov:2021rjp}, we adopt 
the expression obtained by Ref.~\cite{Degrande:2016oan,Khoze:2017tjt,Khoze:2017ifq} 
that the explosive production of $\mathcal{O}(100)$ Higgs bosons leads to the 
factorial growth of scattering amplitude.}. 
The balance between the Higgsplosion and the Higgspersion is nontrivial: 
the annihilation of heavy DM of $\gtrsim \mathcal{O}(1)$~TeV generates an energy 
clump, which transits into $n$-Higgs bosons ($n \gg 1$) before emitting a large 
number of final-state particles. However, as we show below, 
the Higgs multiplicity in this transition is uniquely determined by the value of 
the Higgs self-coupling $\lambda$ by taking the 
thermal average of the DM annihilation cross section. 
Such a mechanism enhances the effective reaction rate between the DM and the SM 
thermal bath, allowing heavier DM compared with those in simple Higgs portal scenarios.

{\it Setup and Formulation---}%
We minimally extend the SM with a fermionic DM $\chi$ which couples to the SM Higgs 
field $\phi$ to demonstrate the Higgsplosion effect on DM relic density. 
The formulation here is applicable to DM scenarios with other scalar fields. 
We introduce the Lagrangian as
\begin{eqnarray}
 \mathcal{L}
   &=& \frac{1}{2}(\partial\phi)^2-\frac{1}{4}\lambda(\phi^2-v^2)^2 \nonumber \\
   & & +\bar{\chi}(i\partial\hspace{-0.5em}/-m_\chi)\chi-\left(y_\chi\phi\overline{\chi_R}\chi_L+({\rm h.c.})\right), 
\label{eq:Lagrangian1}
\end{eqnarray}
where $v$ is the vacuum expectation value of the Higgs field. 
In general, the Yukawa coupling $y_\chi$ is a complex while the DM mass $m_\chi$ 
is chosen to be real by rotating the phase of the left- and right-handed fermion 
fields independently. Redefining the left-handed DM field as 
$e^{i\,{\rm arg}M_\chi}\chi_L \ \longrightarrow \ \chi_L$, 
the DM mass is 
\begin{equation}
M_\chi \ \equiv \ m_\chi+y_\chi v \ = \ |M_\chi|e^{i\,{\rm arg}M_\chi}.
\end{equation}
Then, the Lagrangian (\ref{eq:Lagrangian1}) is represented as
\begin{eqnarray}
 \mathcal{L}
  &=& \frac{1}{2}(\partial h)^2
  -\frac{1}{2}m_h^2h^2-\lambda vh^3
  -\frac{1}{4}\lambda h^4 \nonumber \\
  & & +\bar{\chi}(i\partial\hspace{-0.5em}/-|M_\chi|)\chi 
  -h\bar{\chi}\left(\tilde{y}_\chi P_L+\tilde{y}_\chi^*P_R\right)\chi, 
\label{eq:Lagrangian2}
\end{eqnarray}
where we denote $h=\phi-v$, $m_h^2=2\lambda v^2$, 
$\tilde{y}_\chi \ = \ y_\chi e^{-i\,{\rm arg}M_\chi}$, and $P_{L(R)}$ is the chirality projection 
operator. In the nonrelativistic regime, for real $\tilde{y}_\chi$, 
the annihilation is $p$-wave dominant and suppressed due to low velocity.
On the other hand, the $s$-wave dominant contribution is achieved for complex cases. 
A nonzero value of $m_\chi$ is required for ${\rm Im} \, \tilde{y}_\chi \neq 0$ because 
$\tilde{y}_\chi = y_\chi e^{-i\,{\rm arg}\: y_\chi} = |y_\chi|$ holds for $M_\chi=0$.


\begin{figure}[t!]
 \centering
 \hspace{3em} \includegraphics[keepaspectratio, scale=0.45]{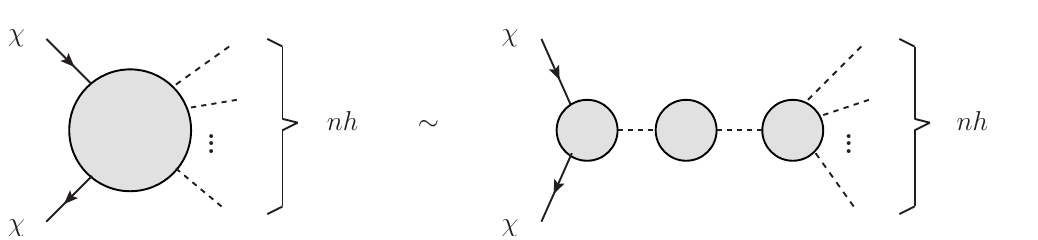}
 \caption{Schematic picture of our approximation in the calculation of the scattering 
 amplitude for $\chi\chi\rightarrow nh$.}
 \label{fig:amp_approx}
\end{figure}

{\it DM annihilation with the Higgsplosion effect---}%
We consider DM that is much heavier than the SM Higgs, $|M_\chi|\sim n m_h$ with 
$n\sim{\cal O}(100)$. We apply the formulation of the Higgsplosion phenomenon 
to the thermally averaged cross section of DM annihilation, $\chi \chi \to nh$. 
The intermediate state in the DM annihilation process is highly excited, and 
its decay rate exponentially grows with the final multiplicity.

We decompose the scattering amplitude for the process $\chi \chi \rightarrow nh$ into three components: 
(i) an effective vertex on DM annihilation $\chi \chi \rightarrow h^*$, 
(ii) a dressed propagator on $h^*\rightarrow h^*$, and 
(iii) an effective vertex on $h^*\rightarrow nh$~ (see FIG.~\ref{fig:amp_approx}). 
This prescription is applicable whenever the $s$-channel type diagram dominates 
over those from $t$- and/or $u$-channel type processes, $\chi \chi \to h^* h^* 
\to n'h + (n-n')h$, where $n'$ is an integer smaller than $n$. We will show later 
that the above situation is generally realized.

With this prescription, the squared amplitude is obtained as 
\begin{widetext}
\begin{eqnarray}
 \sum_{\rm spins}\left|\mathcal{M}(\chi\chi\rightarrow nh)\right|^2
  &\sim& \sum_{\rm spins}
  \left| 
  	\mathcal{M}(\chi \chi \rightarrow h) 
  	\frac{1}{s-M_h(s)^2-iM_h(s)\Gamma_{h}(s)}\mathcal{M} 
  	(h\rightarrow nh)
  \right|^2 \\
  &\sim& 2|\tilde{y}_\chi|^2(s-4|M_\chi|^2\cos^2\theta_{\tilde{y}}) 
  \frac{1}{s^2+m_h^2\Gamma_h(s)^2} 
  \left|\mathcal{M}(h\rightarrow nh)\right|^2, 
\label{eq:amp_chichibar-to-nphi}
\end{eqnarray}
\end{widetext}
where we denote $\theta_{\tilde{y}}\equiv {\rm arg} \: \tilde{y}_\chi$, $M_h(s)$ 
as the effective mass of $h$ including the self-energy correction, and $\Gamma_h(s)$ 
as the total width of the virtual $h$.  
We approximate the effective vertex on $\chi \chi \rightarrow h^*$ by the elementary one in the Lagrangian (\ref{eq:Lagrangian2}), and apply $M_h(s)\sim m_h \ll \sqrt{s}$ in this calculation.

The annihilation cross section associated with all $n$ contributions of the Higgsplosion process 
$\chi \chi \rightarrow nh$ is evaluated using Eq.~(\ref{eq:amp_chichibar-to-nphi}) as
\begin{eqnarray}
\sigma \ = \ \sum_n \sigma_{\chi\chi\rightarrow nh}
  &=& \frac{1}{\sqrt{s(s-4|M_\chi|^2)}}\cdot 
     \frac{|\tilde{y}_\chi|^2}{4} \nonumber \\
  & & \ \times \left(1-\frac{4|M_\chi|^2\cos^2\theta_{\tilde{y}}}{s}\right)W(s) 
\label{eq:total_cross-section_higgsplosion}.
\end{eqnarray}
Here we introduced a ``window'' function $W(s)$
\begin{eqnarray}
    W(s)
     &=& \frac{2sm_h^2\mathcal{R}(s)}{s^2+m_h^4\mathcal{R}(s)^2}, 
     \qquad 0\leq W(s) \leq 1, \label{eq:def_window}
\end{eqnarray}
with the dimensionless reaction rate 
$\mathcal{R}(s) = \Gamma_h(s)/m_h$, which is an exponentially increasing function at 
$\sqrt{s}\gtrsim 100m_h$. See {\it Supplemental Material} for the details.
The window function $W(s)$ consists of the Higgs dressed propagator and the Higgsploding 
vertex on the process $h^*\rightarrow nh$.

\begin{figure}[t]
    \centering
    \includegraphics[scale=0.5]{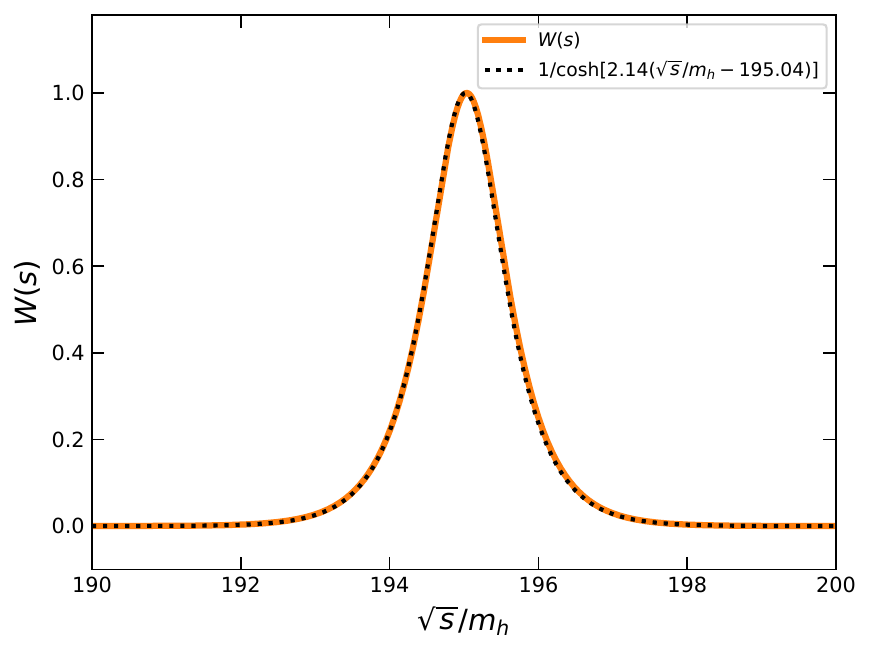}
    \caption{The shape of the window function. The peak position is 
    $\sqrt{s_{\rm peak}}/m_h = 195.04$. The window function can be well-fitted 
    as $W(s)\sim 1/\cosh{\left[2.14(\sqrt{s}-\sqrt{s_{\rm peak}})/m_h\right]}$.
    }
    \label{fig:window_fuction}
\end{figure}

As mentioned above, the reaction rate $\mathcal{R}(s)$ for the $h^*\rightarrow nh$ process 
exponentially grows as $n\sim\sqrt{s}/m_h$ increases. 
Simultaneously, the dressed propagator regulates the exponential growth of $\mathcal{R}(s)$ 
in the denominator of (\ref{eq:def_window}) at large $s$ (Higgspersion effect).
The window function $W(s)$ characterizes the balance between the Higgsplosion and the Higgspersion: 
a smaller or larger $\mathcal{R}(s)$ provides a suppression factor in 
(\ref{eq:total_cross-section_higgsplosion}), while the window
opens ($W(s)\sim 1$) at a certain point $s=s_{\rm peak}$ 
satisfying the peak condition
$s_{\rm peak}=m_h^2\mathcal{R}(s_{\rm peak})$. 
Both the peak position and the shape of the window function $W(s)$ depend only on the Higgs 
self-coupling $\lambda$. Using $\lambda=0.129$, the peak appears at 
$\sqrt{s_{\rm peak}} \simeq 195 m_h$ with a finite width of 
$\Delta \sqrt{s} \simeq \pm 1 m_h$ (see FIG.~\ref{fig:window_fuction}). 
This means that the process $\chi \chi \rightarrow 195h$ has the leading contribution 
among the annihilation channels to various $n$.

The structure of the window in the $s$-channel type process explains the reason 
this process dominates over the $t$- or $ u$-channel types. 
The contribution from $t$- or $u$-channel type processes is regarded as the double 
Higgsplosion process of $\chi+\chi\to h^*+h^*\to n'h+(n-n')h$. If the multiplicity 
of the two Higgs in the intermediate process is similar ($n'\sim n/2$), then the 
spectrum cannot fit the window. On the other hand, if one of the intermediate-state 
Higgs takes most of the multiplicity ($n'\ll n$), then the process is that of a 
Higgsplosion with an additional external Higgs; hence the additional suppression 
factor $\left(\tilde{y}_\chi/\sqrt{4\pi}\right)^{n'}$ compared to the $s$-type 
process is expected~\footnote{The summnation over all possible $n'$ would result in 
non-trivial outcomes. A careful investigation incorporating the non-perturbative 
effects which are originated from $t$-/$u$-channel type processes is required.}.

Here we emphasize the important role of the phase of the coupling $\tilde{y}_\chi$. 
As seen in Eq.~(\ref{eq:total_cross-section_higgsplosion}), a pure imaginary 
$\tilde{y}_\chi$ ($\theta_{\tilde{y}}= \pm\pi/2$) leads to the $s$-wave annihilation, 
while a real $\tilde{y}_\chi$ ($\theta_{\tilde{y}}=0$) leads to the $p$-wave feature  because 
the cross section is proportional to the momentum of $\chi$, 
$\sqrt{s-4|M_\chi|^2} \sim |\vec{p}_\chi|$. 
The cross section of the DM pair-annihilation into two Higgs bosons serves as a 
good reference. It is evaluated as 
\begin{widetext}    
\begin{equation}
    \sigma(\chi\chi\rightarrow hh) 
    \ \sim \ 
    \frac{1}{\sqrt{s(s-4|M_\chi|^2)}} \cdot \frac{|\tilde{y}_\chi|^4}{16\pi} 
    \left[ 
    \sin^2 2\theta_{\tilde{y}} 
    + \frac{1}{3}(4\cos^2\theta_{\tilde{y}}-1)^2 
    \left(1-\frac{4|M_\chi|^2}{s}\right)+\cdots 
    \right] 
\label{eq:total_cross-section_2-to-2}
\end{equation}
\end{widetext}
in the limit of $\sqrt{s}\gg m_h$.  The first and second terms inside the brackets correspond 
to the $s$- and $p$-wave contributions, respectively. 
As is shown in 
Eqs.~\eqref{eq:total_cross-section_2-to-2} and \eqref{eq:total_cross-section_higgsplosion}, 
at low-momentum regimes, the pure imaginary $\tilde{y}_\chi$ damps the 2-to-2 annihilation, 
and while it maximizes the 2-to-$n$ annihilation processes.


{\it Boltzmann equation and numerical results---}%
The cross-section incorporating the Higgsplosion is now applied to the evaluation of thermal 
relic abundance. The Boltzmann equation for the DM number density is 
\begin{eqnarray}
     \dot{n}_\chi+3Hn_\chi &=& -\langle\sigma v\rangle\left(n_\chi^2-(n_\chi^{\rm eq})^2\right), 
     \label{eq:Boltzmann_eq}
\end{eqnarray}
where $H$ is the Hubble parameter. $n_\chi$ and $n_\chi^{\rm eq}$ are the actual and 
equilibrium number density of DM, respectively. Here, we assume the kinetic equilibrium 
between DM and background SM fields. The thermally-averaged cross section is derived using 
Eq.~(\ref{eq:total_cross-section_higgsplosion}) as 
\begin{eqnarray}
    \langle\sigma v\rangle
     &=& \frac{1}{(n_\chi^{\rm eq})^2}\cdot 
     \frac{|\tilde{y}_\chi|^2}{32\pi^4}T^4\int_{4|M_\chi|^2}^\infty 
     \frac{ds}{s} \rho(\sqrt{s}/T) \: W(s),  
\label{eq:thermal_ave_cs}
\end{eqnarray}
where $T$ is the temperature and
\begin{equation}
    \rho(\sqrt{s}/T) \ 
    = \ \frac{\sqrt{s-4|M_\chi|^2}\left(s-4|M_\chi|^2\cos^2\theta_{\tilde{y}}\right)}{T^3} K_1(\sqrt{s}/T) \label{eq:scattering_spectrum}
\end{equation}
describes the spectrum of the scattering.  $K_n(x)$ is the $n$-th modified Bessel function.

\begin{figure}[t]
    \centering
    \includegraphics[scale=0.42]{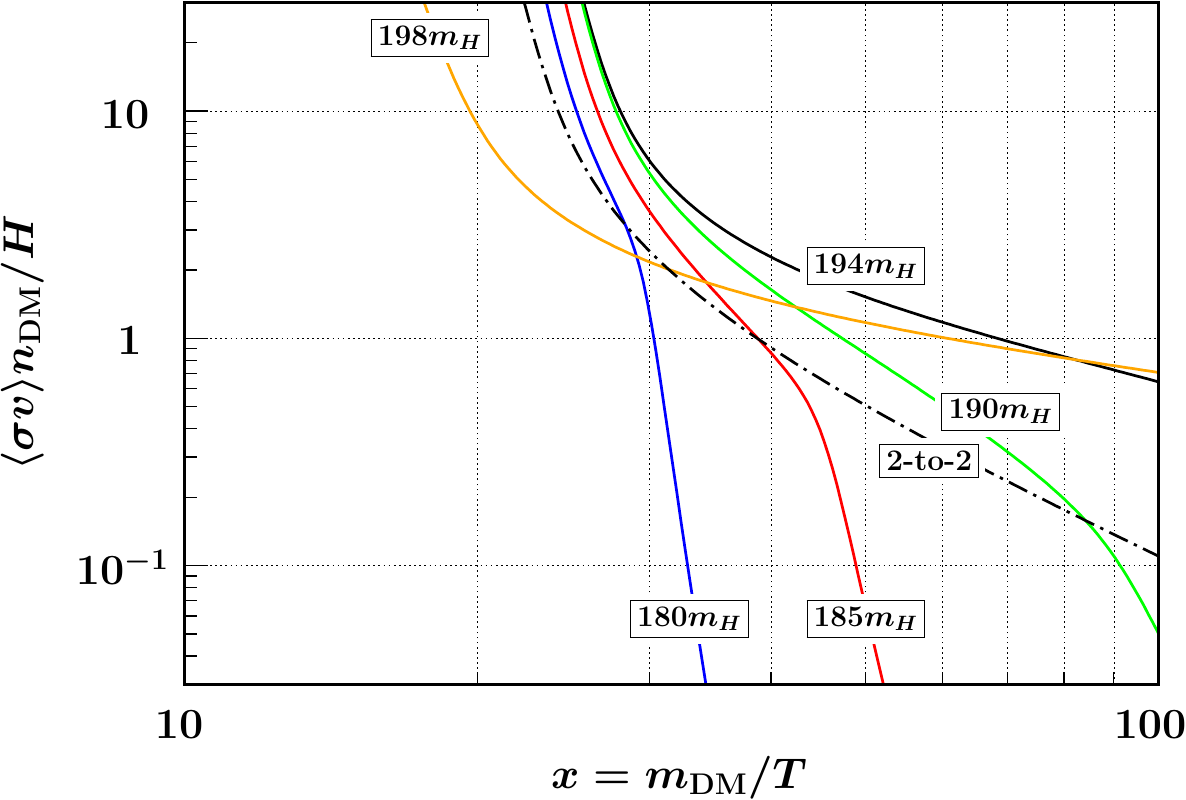}
    \includegraphics[scale=0.42]{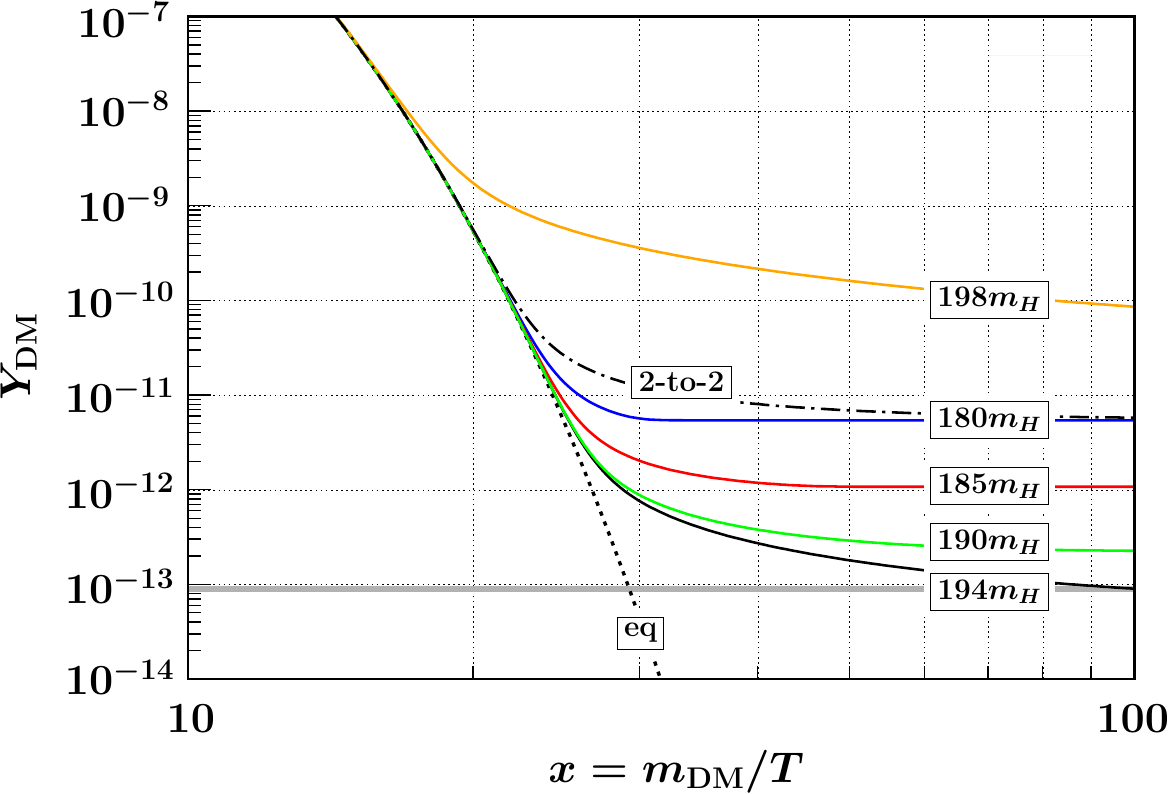}
    \caption{Evolution of the ratio of reaction rates to the Hubble parameter 
    (upper panel) and the yield of DM (bottom panel) for each DM mass.  
    Parameters are set as $\lambda=0.129$, $m_h=50$ GeV, and $\tilde{y}_\chi=1.30i$. 
    The results by the 2-to-2 scattering process ($\chi \chi \rightarrow 2h$) are 
    evaluated based on (\ref{eq:total_cross-section_2-to-2}) with 
    $2|M_\chi|=194m_h=2\times 4.85$ TeV.}
    \label{fig:abundance}
\end{figure}

FIG.~\ref{fig:abundance} shows the numerical results with $\lambda=0.129$, 
$m_h=50$ GeV and $\tilde{y}_\chi=1.30i$. 
Each line corresponds to a different DM mass case. 
The reason for the small Higgs mass $m_h=50$ GeV $< 125$ GeV is that the 
freeze-out era is expected in electroweak symmetry breaking. 
The freeze-out temperature of the DM annihilation can be estimated as 
$T_f\sim |M_\chi|/25 \sim \frac{1}{2}nm_h/25\sim 200$ GeV $\times (m_h / 50 {\rm GeV})$ 
for $n\sim 200$. This temperature must be lower than the electroweak symmetry breaking 
scale, since otherwise the Higgs particle is massless.

In the upper panel of FIG.~\ref{fig:abundance}, the ratio of the reaction rate 
$n_\chi \langle \sigma v \rangle$ to the Hubble parameter $H$ is shown. 
The annihilation process freezes out when the ratio reaches the order of unity. 
As shown in the figure, the maximum reaction rate is achieved at
$2|M_\chi| \sim 194m_h$, which is slightly smaller than the window peak 
$\sqrt{s_{\rm peak}}=195m_h$.
For heavier DM cases of $2|M_\chi| \gtrsim 194m_h$, the scattering spectrum in 
(\ref{eq:thermal_ave_cs}) is always out of the window since $s > \sqrt{s_{\rm peak}}$. 
At this regime, its cross section damps.
On the other hand, for cases of $2|M_\chi| \lesssim 194m_h$, although the window function 
still opens, the dominant contribution in the cross-section spectrum 
(\ref{eq:scattering_spectrum}) does not match the window range due to the Boltzmann 
suppression in $K_1(\sqrt{s}/T)$.
Consequently, the DM with $2|M_\chi| \sim 194m_h$ can be in the chemical 
equilibrium for the longest time and reduce the DM yield sufficiently.

In the lower panel, we show the DM yield $Y_\chi=n_\chi/s$, where 
$s=\frac{2\pi^2}{45}g_{*S} T^3$ is the entropy density. Here $g_{*S}$ denotes the 
total number of relativistic degrees of freedom. 
The current observation requires~$Y_{\chi, {\rm now}}=8.99 \times 10^{-14} 
\left(4.85 \ {\rm TeV}/|M_\chi| \right)$~\cite{Planck:2018nkj}. 
Our calculation finds $Y_\chi(x=100)=9.08\times 10^{-14}$ for 
$2|M_\chi|=194m_h=2\times 4.8$~TeV, and hence this representative 
parameter set successfully accounts for the relic abundance.

We here comment on the smaller reaction rates than that of the 2-to-2 annihilation 
channel in FIG.~\ref{fig:abundance}, the results for $2|M_\chi|=180m_h$ and $185m_h$. 
This unexpected result originates from applying the dimensionless reaction rate 
$\mathcal{R} (s)$ to the 2-to-2 process. The $\mathcal{R} (s)$ is computed in the 
non-relativistic limit of the final state particles. 
For low-multiplicity production processes, this limit is no longer applicable. 
In our computation of the total cross section, however, all channels of $n$-body 
final states are summed over by applying the 
$\mathcal{R} (s)$ regardless of the limited applicability. 
The contributions from low-multiplicity processes are small compared 
to those of high-multiplicities. (See {\it Supplemental material} for details.)
It is rational and acceptable for the purpose of highlighting the feature of 
the Higgsplosion. The complete Higgsploding cross section, as a practical matter, 
is always larger than the case of the 2-to-2 process.

{\it Discussion---}
This work highlights the impact of the Higgssplosion phenomenon on the relic 
abundance of Higgs-portal DM, taking a simplified setup of a minimally extended model. 
It can drastically change the DM abundance in Higgs-portal DM models, but further 
investigations are needed to obtain a conclusive picture. There are several issues to 
be investigated.  
For example, quantum statistical effects for the high-multiplicity state would change 
the evolution of the relic density by distorting the shape of the window function. 
Further enhancement of the thermally-averaged cross section through the stimulated 
emission would be expected if the Bose-Einstein statistics is applied. A careful 
formulation of the intermediate dressed state, where the correlation function is 
regulated by thermal damping and Debye mass, which is beyond the scope of this work.

The signatures in the current Universe should be carefully searched for. In the model we 
discussed here, the DM annihilation to high-multiplicity final states cannot be expected 
in the current Universe. This is because $m_h=125$~GeV has already been achieved, and the 
kinematically allowed Higgs-multiplicity is $2|M_\chi|/m_h^{T=0} \sim 77$, which is 
outside the window function. Therefore, the energetic 2- or 3-body Higgs final state is 
the leading DM annihilation channel. 
From Eq.~\eqref{eq:total_cross-section_2-to-2}, the maximal annihilation cross section is 
$\sigma v(\chi \chi \to 2 h) \simeq |\tilde{y}_\chi|^4/32 \pi M_\chi^2$, and is numerically 
$1.41 \times 10^{-26} \, \text{cm}^3 \text{s}^{-1}$ for $|\tilde{y}_\chi|=1.30$ and $M_\chi 
= 4.85\,$TeV. This is slightly smaller than the current bounds from indirect detection 
experiments $\sigma v|_{\text{obs.}} \lesssim 5 \times 10^{-25} 
\, \text{cm}^3 \text{s}^{-1}$ for $M_\chi = 4.85\,$TeV~\cite{Calore:2022stf}. Future 
observation (e.g.~\cite{LHAASO:2019qtb,Albert:2019afb}) will be able to probe the signature. 
Null results from direct searches for DM ruled out some of the parameters that characterize 
DM. An idea to evade direct detection bounds is to make the DM a Majorana state~\cite{Basirnia:2016szw}. 
It is realized by considering a minimal extension with a singlet Majorana fermion $\Psi$ and two kinds of 
$SU(2)_L$ fermions: $f_L (= (f^-,f^0)^T)$ 
and $f_R (=(f'^-,f'^0)^T)$ of which the Lagrangian includes terms 
    $-\frac{1}{2} \bar{\Psi} \left(m_\psi P_L+m_\psi^*P_R\right) \Psi 
    -m_D \bar{f} f
    -\left(\bar{\Psi}\left(y_L P_L+y_R P_R\right)H \cdot f
    +h.c.\right)$. 
We take positive $m_D$ and $y_{L,R}$ without losing generality, while $m_\psi$ is a complex. 
In the mass eigenbasis, the interactions of the lightest state $\chi_1$, which is DM, with 
the SM Higgs and the $Z$ boson are parametrized as $ h \bar{\chi}_1(c_h+ic_5\gamma^5) \chi_1$ 
and $c_Z Z_\mu \bar{\chi}_1 \gamma^\mu \gamma^5 \chi_1$. The coefficients $c_h$ and $c_Z$ 
are constrained from spin-independent and spin-dependent direct detection 
experiments~(e.g.~\cite{Basirnia:2016szw}). 
For typical scales expected from the DM relic abundance in our setup, e.g., 
$m_\psi =i\times 5$\,TeV and $m_L = m_{\bar{L}} = 1$\,TeV, then 
$|c_h| \simeq \frac{y_L y_R v}{m_D}\frac{{\rm Re} \, m_\psi}{|m_\psi|} = 0$ and 
$|c_Z| \simeq \frac{g_2}{4\cos\theta_w}\frac{|y_L^2-y_R^2|v^2}{2m_D^2} = 0$ in the 
mostly-singlet limit, where $|m_D|\gg |m_\psi|,\ |m_L|,\ |m_{\bar{L}}|$ and $y_L=y_R$. The 
remaining coupling, $c_5$, is difficult to constrain with current facilities~\cite{Goodman:2010yf}.

Our results are applicable to other scenarios, where the DM annihilates into 
neutral scalars. The important ingredient is the window function $W(s)$; the self-coupling 
constant of neutral scalars shapes it, and its width and peak position govern the 
final state multiplicity. 
A highlighted difference between the DM scenarios with general scalars and SM 
Higgs is the final state multiplicity of DM annihilation in the current Universe. 
As was mentioned, since $m_H^{T=0} \neq m_H^{T=T_f}$ for the DM of $M_\chi 
\simeq 4.85$\,TeV, no explosion happens in the DM annihilation at the present day. 
On the other hand, if $m_h^{T=0} = m_h^{T=T_f}$, the high-multiplicity final state 
channel would be leading in the DM annihilation either at present or the 
freeze-out epoch. 
It is important to emphasize here that the DM annihilation receives 
quantum statistical effects (e.g., stimulated emission) in the early Universe but not 
in the current Universe. It might lead to a significant discrepancy between these reaction rates.  
The relic density and indirect signals of scalar-portal DM scenarios should be 
carefully improved by taking into account the high-multiplicity final state with 
statistical corrections. 
These issues will also be investigated in future work.

{\it Summary---}
Cosmological consequences of the Higgsplosion effect revise our picture of DM. 
In this {\it Letter}, we revisit the relic abundance of Higgs-portal DM considering 
the high multiplicity final state due to the Higgsplosion. 
A key ingredient for the evolution of DM density is the 
window function~(see Eq.~\ref{eq:def_window}), which is obtained by carrying out 
the thermal average of DM 
annihilation cross section, considering contributions from both the Higgsplosion 
and the Higgspersion effects. 
The Higgsplosion stands for the factorial growth of final-state multiplicity in the 
transition $h^* \to nh$, which maintains DM longer in the thermal bath of the SM fields. 
This high multiplicity transition also makes a major correction to the regulator 
of the energetic intermediate Higgs, which retains the unitarity in this transition, referred to 
as the Higgspersion. The window depends only on the Higgs self-coupling; the peak 
position is $\sqrt{s}\sim 195 m_h$ with a finite width of $\Delta \sqrt{s} \sim \pm 1m_h$ for 
$\lambda \simeq 0.129$. 
In a minimal Higgs-portal DM model with these effects, DM of 
$M_\chi\simeq 4.85$~TeV successfully accounts for the observed relic abundance when the 
complex coupling to Higgs $\tilde{y}_{\chi}\sim 1.30i$, without conflicting with either of the 
direct or indirect constraints.
Our result is applicable to a wider class of models with other scalar fields,  
opening a new window for heavy DM.

{\it Acknowledgments---}%
The authors thank T. Nomura for the helpful discussion. 
This work is supported in part by the Scientific Research (22K14035 [NH], 22K03638, 22K03602 [MY], 20H05852 [NH,KM,KY]).
The works of NH and SE are also supported in part by the MEXT Leading Initiative for Excellent Young Researchers Grant Number 2023L0013.
KM is supported by NSF Grants Nos.~AST-2108466, AST-2108467, and AST-2308021. 
This work was partly supported by MEXT Joint Usage/Research Center on 
Mathematics and Theoretical Physics JPMXP0619217849 [MY].

\noindent\rule{8.5cm}{0.3pt}
\section*{Supplemental Material}
\begin{center} 
Explosive production of Higgs particles and implications for heavy dark matter: calculation of the thermally-averaged cross-section
\end{center}
\noindent\rule{8.5cm}{0.3pt}

\begin{figure}[h]
    \centering
    \includegraphics[scale=0.45]{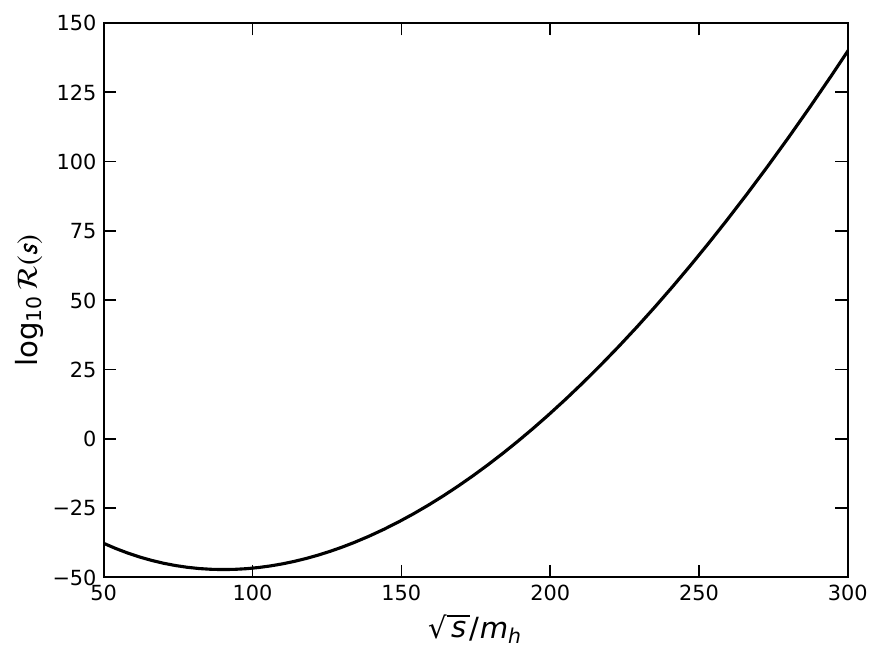}
    \caption{The plot of the dimensionless reaction rate as a variable $\sqrt{s}/m_h$.}
\label{fig:logR}
\end{figure}

In this {\it Supplemental Material}, we discuss the details of the cross section through the Higgsploding process and its thermally-averaged quantity. We begin with the formula of the cross section given as Eq.~(6) in the main text:
\begin{eqnarray}
\sigma \ &=& \ \sum_n \sigma_{\chi\chi\rightarrow nh}
  \\
  &=& \frac{1}{\sqrt{s(s-4|M_\chi|^2)}}
  \cdot \frac{|\tilde{y}_\chi|^2}{4} \left(1-\frac{4|M_\chi|^2\cos^2\theta_{\tilde{y}}}{s}\right)W(s)
\end{eqnarray}
where the window function $0\leq W(s)\leq 1$ is given by
\begin{eqnarray}
    W(s)
     &=& \frac{2sm_h^2\mathcal{R}(s)}{s^2+m_h^4\mathcal{R}(s)^2} \ 
     = \ \frac{1}{\cosh[\ln(\mathcal{R}(s)m_h^2/s)]} \label{eq:def_window}
\end{eqnarray}
with the dimensionless reaction rate 
\begin{widetext}
\begin{eqnarray}
 \mathcal{R}(s)
  &=& 
  \sum_n\theta(\sqrt{s}-nm_h)\mathcal{R}_n(s), \\
  \mathcal{R}_n(s)
  &=& \frac{1}{2m_h^2}\frac{1}{n!} 
  \int\frac{d^3p_{h_1}}{(2\pi)^3} \cdots 
  \frac{d^3p_{h_n}}{(2\pi)^3} 
  \frac{1}{2E_{h_1}\cdots 2E_{h_n}} (2\pi)^4\delta^4(p_h-p_{h_1}-\cdots p_{h_n}) 
  \left|\mathcal{M}(h\rightarrow nh)\right|^2 \nonumber \\
  &\sim& \exp\left[
  n\left( \frac{2}{\sqrt{3}}\frac{\Gamma(5/4)}{\Gamma(3/4)}\sqrt{\lambda n} +\ln\frac{\lambda n}{4e} 
  +\frac{3}{2}\ln\left(\frac{e}{3\pi}\frac{\sqrt{s}-nm_h}{nm_h}\right)
  -\frac{25}{12}\frac{\sqrt{s}-nm_h}{nm_h}\right)
  \right]
\end{eqnarray}
\end{widetext}
introduced in Ref.~\cite{Khoze:2017tjt}. Note that 
the complete picture including the nonperturbative effects is still a subject of ongoing discussions.  
The dimensionless reaction rate $\mathcal{R}(s)$ increases exponentially at high $s$ as 
shown in FIG.~\ref{fig:logR}.
Applying the above formulae to Eq.~(\ref{eq:def_window}), we obtain the shape of the window 
function as seen in FIG.~2 of the main text, which leads to the peak position at 
$\sqrt{s_{\rm peak}}=195.02m_h$. The dependence on the final-state Higgs multiplicity directly 
reflects the properties of the window function $W(s)$.

\begin{figure}[h]
    \centering
    \includegraphics[scale=0.42]{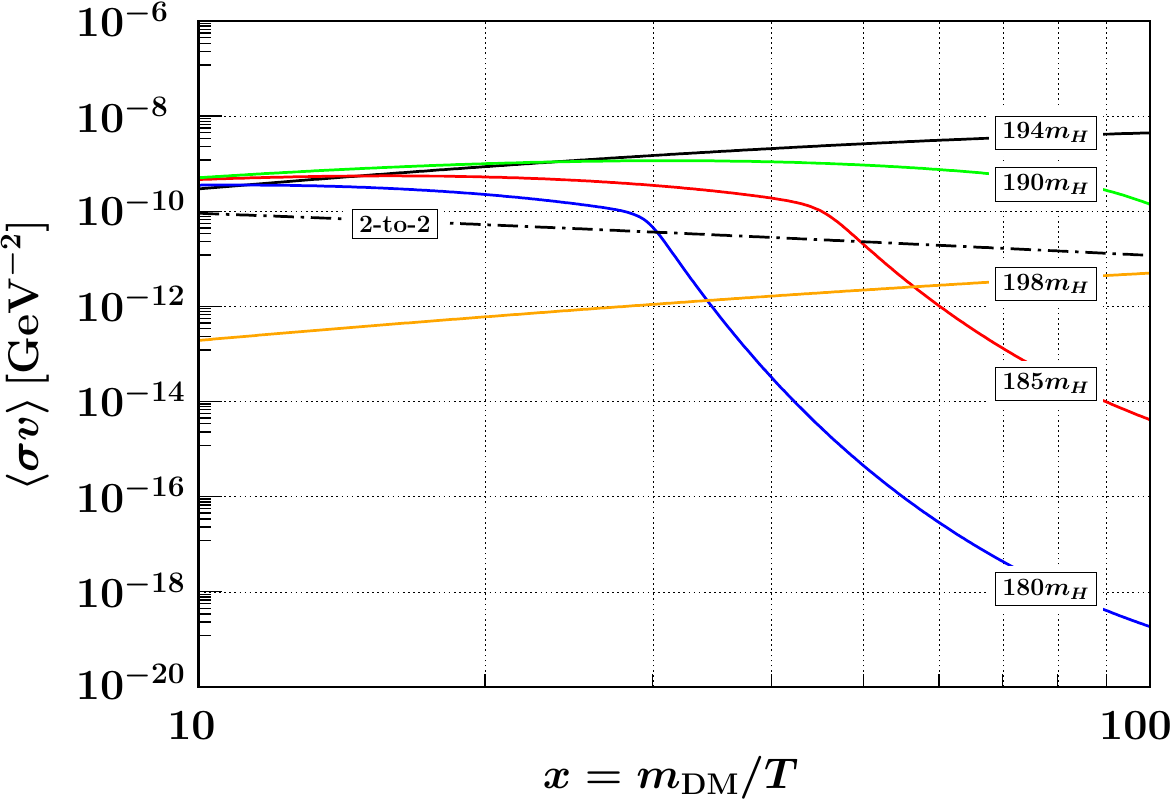}
    \caption{Evolution of the thermally-averaged cross section for 
    each DM mass. Parameters are set as $\lambda=0.129$, 
    $m_h=50$~GeV, and $\tilde{y}_\chi=1.30i$. The results by 
    the 2-to-2 scattering process is evaluated with 
    $2|M_\chi|=194m_h=2 \times 4.85\,\text{TeV}$. }
\label{fig:TAXS}
\end{figure}

The thermally-averaged cross section given as 
Eq.~(13) in the main text:
\begin{eqnarray}
    \langle\sigma v\rangle
     &=& 
     \frac{1}{(n_\chi^{\rm eq})^2}\cdot \frac{|\tilde{y}_\chi|^2}{32\pi^4}T^4 
     \int_{4|M_\chi|^2}^\infty \frac{ds}{s} 
     \rho(\sqrt{s}/T) \: W(s),  
\label{eq:thermal_ave_cs}
\end{eqnarray}
where $\rho$ is the scattering spectrum, and $W$ stands for the window function. 
FIG.~\ref{fig:TAXS} shows the behavior of $\langle\sigma v\rangle$ varying the number of Higgs bosons produced from the process $\chi\chi\to nh$ with $n\gg2$.
As seen in FIG.~\ref{fig:TAXS}, the thermally-averaged cross section continuously 
grows with $x=M_\chi/T$ for $2m_\chi=194 m_h$ and $198 m_h$ while the others show 
drops at large $x$, i.e. low $T$. 

Eq.~(\ref{eq:thermal_ave_cs}) with an approximated 
estimation of $W(s)$ helps to understand the above behaviors in this context. 
Expanding the function $\ln\left(R(s)m_h^2/s\right)$ around $\sqrt{s}=\sqrt{s_{\rm peak}}$ 
with a condition $\mathcal{R}(s_{\rm peak})m_h^2/s_{\rm peak}=1$, one  obtains
\begin{equation}
\begin{split}
    W(s) 
    &\simeq \frac{1}{\cosh \left[ \alpha \frac{\sqrt{s}-\sqrt{s_{\text{peak}}}}{m_h} \right]} 
    \\&
    \simeq \frac{1}{2} \exp \left[ -\alpha \left( N-N_* + \frac{Ny}{2x} \right) \right],
    \label{eq:ws0} 
\end{split}     
\end{equation}
where $\alpha \equiv \left.\left(\frac{m_h}{\mathcal{R}}\frac{d\mathcal{R}}
{d\sqrt{s}}-\frac{2m_h}{\sqrt{s}}\right)\right|_{s=s_{\rm peak}} \simeq 2.14$. 
It is useful to scale out with the redefinitions of a constant of 
$N=2M_\chi/m_h$ 
and an integral variable $y$ by $\sqrt{s}=2M_\chi+Ty$ to obtain the approximated 
formula of the second line of Eq.~\ref{eq:ws0}. $N_*$ corresponds to $N$ at 
$s_{\rm peak}$, i.e. $N_*=\sqrt{s_{\rm peak}}/M_\chi\sim 195$ and we take 
$\theta_{\tilde{y}}=[{\rm arg} M_\chi] =\pi/2$ to maximize the thermally-averaged 
cross section. Finally, the thermally averaged cross section is obtained as
\begin{equation}
    \langle \sigma v \rangle 
    \propto 
    \int_0^\infty dy \sqrt{y} \, e^{-y} 
    \exp\left[ -\alpha \left( N-N_* + \frac{Ny}{2x} \right) \right].
    \label{eq:ws} 
\end{equation}

Two sources of the exponential suppression appear in Eq.~\ref{eq:ws}: 
the first one with $e^{-y}$ originates from the Boltzmann suppression, which 
effectively restricts the range of $y$-integral within $0\leq y \lesssim 1$.
The latter is the suppression unique to this model which incorporates the Higgsplosion 
mechanism. For $N \sim 200\gtrsim N_*$, the factor $\exp[-\alpha Ny/2x]$ dominates 
over $\exp[-y]$ up to $x \lesssim \frac{1}{2}\alpha N \sim 214$, and hence this 
integrant grows in the regime $10 \leq x\leq 10^2$ which we show in the figure.
For much larger values of $x\gg 214$, the term with $\exp[-\alpha Ny/2x]$ 
goes to the unity then the thermally-averaged cross section drops by the factor 
$e^{-\alpha (N-N_*)}$. 
On the other hand, the damping feature at the lower multiplicity $N\lesssim N_*$ 
has a different reason. In the early stage of $x\lesssim{\cal O}(10)$, the 
``effective'' integration range $0\leq y \lesssim 1$ covers the peak position of 
the window enough, while the effective 
range cannot reach the peak of the window at the later stage $x\gtrsim{\cal O}(10)$.  
This means that the kinetic energy of 
the annihilating DM particles is insufficient to induce the Higgsplosion.

\bibliographystyle{unsrtnat}
\bibliography{references}

\end{document}